\documentclass[a4paper,10pt]{article}
\usepackage[utf8]{inputenc}
\usepackage{hyperref}
\usepackage{epsfig}
\usepackage{amsmath}
\usepackage{amssymb}
\usepackage{times}
\usepackage{indentfirst}
\usepackage{multicol}
\usepackage{color}
\usepackage{xcolor,graphicx}
\usepackage{moreverb}
\usepackage[top=1.0in,bottom=1.0in,left=1.25in,right=1.25in]{geometry}
\usepackage{wrapfig} 
\usepackage{authblk}
\title{Stock market as temporal network}
\author[1,4]{Longfeng Zhao\thanks{zlfccnu@mails.ccnu.edu.cn}}
\author[2]{Gang-Jin Wang}
\author[3,4]{Mingang Wang}
\author[5]{Weiqi Bao}
\author[1]{Wei Li\thanks{liw@mail.ccnu.edu.cn}}
\author[4]{H. Eugene Stanley}
\affil[1]{Complexity Science Center and Institute of Particle Physics, Central
	China Normal University, Wuhan 430079, China}
\affil[2]{Business School and Center for Finance and Investment Management, Hunan
	University, Changsha 410082, China}
\affil[3]{School of Mathematical Science, Nanjing Normal University, Nanjing 210042, Jiangsu, China}
\affil[4]{Center for Polymer Studies and Department of Physics, Boston University, Boston,
	MA 02215, USA}
\affil[5]{Commercial College, Shandong University, Weihai, Weihai 264209, China}

\graphicspath{{./}}
\begin{document}

\maketitle

\begin{abstract}
Financial networks have become extremely useful in characterizing the structure of complex financial
systems. Meanwhile, the time evolution property of the stock markets can be described by temporal networks. We utilize
the temporal network framework to characterize the time-evolving correlation-based networks of stock markets.
The market instability can be detected by the evolution of the topology structure of the financial networks.
We employ the temporal centrality as a portfolio selection tool. Those portfolios, which are composed of peripheral stocks with low temporal centrality scores, have consistently better performance under different portfolio 
optimization schemes, suggesting that the temporal centrality measure can be used as new portfolio optimization and risk 
management tools. Our results reveal the importance of the temporal attributes of the stock markets, which should be taken serious consideration in real life applications.
\end{abstract}
 \section{Introduction}
 The correlation-based network has become an effective tool to investigate the correlation between complex financial systems\cite{Marti,Kwapien2012}.
 Different methods have been proposed to probe the complex correlation structure of financial system including the threshold method, the minimum spanning tree(MST)\cite{Mantegna1999a}, the planar maximumly filtered graph(PMFG)\cite{Tumminello2005}
 and a strand of other methods\cite{Kenett2010,csm,Gao2015,Gan2015,WangGJ2017a,WangGJ2017,WangGJ2016}. The common aim of all correlation-based networks is seeking for a sparse representation of the high dimensional correlation matrix of the 
 complex financial system. Unlike other eigenvector-based methods(e.g., the principal component analysis) which decompose the variance of the system into a few dimensions, the 
 correlation-based methods directly map the dense correlation matrix into sparse representation. The easy implementations and straightforward interpretations
 of those methods make them quite popular in complex system analysis, especially for complex financial systems. Recently, the correlation-based network has been used for
 portfolio selection in which some risk diversified portfolios are constructed based on a hybrid centrality measure of the MST and PMFG
 networks of the stock return time series\cite{Pozzi2013}. It is well known that the financial system has its own temporal properties which makes it extremely hard or even 
 impossible to forecast. Thus if we want to construct our portfolio in a proper way, we have to consider the temporal attribute of the financial system. 
 
 In this work, we
 analyze the correlation-based networks of stock markets by using the temporal network paradigm. Specifically we have analyzed the temporal evolution of three major stock markets
 of the world, namely, the US, the UK and China. Based on a centrality measure of temporal network, we also construct some portfolios that consistently perform the best under two 
 portfolio optimization schemes. Our work is the first research that incorporates the temporal network methods into the study of complex financial system. The temporal
 evolution of the topological structures can be used to access the information of market instability. The effectiveness of the temporal centrality measure in portfolio selection
 depicts the importance of the temporal structure for the analysis of stock market. The remainder of the paper is organized as follows: Section \ref{data} gives the data
  description and the methodology we use through the paper. Section \ref{results} presents the main results of the paper including the topology analysis of the stock markets and
  the application to the portfolio optimization problems. Section \ref{conclusion} is the conclusion.
 
 \section{Data and methodology\label{data}}
  \subsection{Data}
 Our data sets include the daily returns of the constitute stocks of three major indexes in the 
 world: S\&P 500 (the US), FTSE 350 (the UK) and SSE 380 (China). After removing those stocks with very small sample size, we still have 
 401, 264, and 295 stocks for the three markets
 respectively. In the S\&P 500 dataset, each stock includes 4025 daily returns from 4 January 1999 to 31 December 2014. The FTSE 350 stocks include 3000 daily returns in the period 
 between 10 October 2005 and 26 April 2017. The SSE 380 stocks consist of 2700 daily returns from 21 May 2004 to 19 November 2014.
 
 \subsection{Cross-correlation between stocks}
 We adopt the logarithm return defined as
 \begin{equation}
 r_{i}(t)=\mathrm{ln}p_{i}(t+1)-\mathrm{ln}p_{i}(t),
 \end{equation}
 \noindent
 where $p_{i}(t)$ is the adjusted closure price of stock $i$ at time $t$. We then
 compute the cross-correlation coefficients between any pair of return time series
 at time $t$ by using the past return records sampled from a moving window with length $\Delta$. We then calculate the similarity between stocks $i$ and $j$ at time $t$ with the traditional Pearson correlation coefficient,
 
 \begin{equation}
 \rho_{ij}^{t,\Delta}=\frac{\langle R_{i}^{t}R_{j}^{t}\rangle-\langle R_{i}^{t}\rangle \langle R_{j}^{t}\rangle}{\sqrt{\left[ \langle  R_{i}^{t^{2}}\rangle -\langle R_{i}^{t}\rangle^{2}\right] \left[ \langle  R_{j}^{t^{2}}\rangle -\langle R_{j}^{t}\rangle^{2}\right] }},
 \end{equation}
 \noindent
 where $\Delta$ is the moving window length, and $\langle\ldots\rangle$ is the sample mean over co-trading days of stocks $i$ and $j$ in
 the logarithm return series vector $R_i^t=\{r_i(t)\}$ and $R_j^t=\{r_j(t)\}$.  We obtain an
 $N\times N$ matrix $\mathrm{\textbf{C}}^{t,\Delta}$ at time $t$ with estimation windows $\Delta$ days, and $N$ is the number of stocks.
 The entries of $\mathrm{\textbf{C}}^{t,\Delta}$
 are cross-correlation coefficients $\rho_{ij}^{t,\Delta}$ between all pairs of stocks. The moving 
 window widths are $\Delta=500$ days for S\&P 500 and $\Delta=300$ days for both FTSE 350 and SSE 380. The moving window widths are chosen to 
 make the correlation matrix non-singular(with $\Delta\geq N$). With moving window width $\Delta$, we shift the moving window with 25 days step, thus we obtain a strand of correlation matrices 
 for three markets. Finally we have 142 correlation matrices for S\&P 500, 109 correlation matrices for FTSE 350 and 97 correlation matrices for SSE 380 respectively.
 \subsection{PMFG network of stock market}
 Since the dense representation given by the cross-correlation matrix will induce lots of redundant information,
 thus it is very hard to discriminate the important information from noise. 
 Here we employ the the planar maximally filtered graph(PMFG) method \cite{Tumminello2005} to
 construct sparse networks based on correlation matrices $\mathrm{\textbf{C}^{t,\Delta}}$. The algorithm
 is implemented as follows,\\
 (i) Sort all of the $\rho_{ij}^{t,\Delta}$ in descending order in an ordered list $l_{sort}$.\\
 (ii) Add an edge between nodes $i$ and $j$ according to the order in $l_{sort}$ if and only 
 if the graph remains planar after the edge is added.\\
 (iii) Repeat the second step until all elements in $l_{sort}$ are used up.
 
 Finally a planar graph $G^{t,\Delta}$ is formed with $N_e=3(N-2)$
 edges. It has been addressed in Ref.\cite{Tumminello2005} that the PMFG not only keeps the hierarchical
 organization of the MST but also induces cliques. We calculate such basic topological quantities as the clustering coefficient $C$ and the shortest-path length
 $L$\cite{albert2002}. A heterogeneity index $\gamma$ \cite{estrada2010quantifying} is also used to measure the heterogeneity of PMFGs which is defined by
 \begin{equation}
 \gamma=\frac{N-2\sum\limits_{ij\in {\{e\}}}(k_{i}k_{j})^{-1/2}}{N-2\sqrt{N-1}},
 \label{heterogenity}
 \end{equation}
 where $k_i$ and $k_j$ are the degrees of nodes $i$ and $j$ connected by edge $\{e_{ij}\}$. We also utilize the Jaccard index\cite{Jaccard1912} $J$ 
 to show the variability of the network structure form $t$ to $t+1$. The Jaccard index $J_{G_1 G_2}$ between networks $G_1$ and $G_2$ is defined as
 \begin{align*}
 J_{G_1,G_2}=\frac{E_{G_1}\cap E_{G_2}}{E_{G_1}\cup E_{G_2}},
 \end{align*}
 where $E_{G_1}$ and $E_{G_2}$ are the edges of networks $G_1$ and $G_2$, respectively.
 \subsection{Supra-Evolution matrix for temporal stock network}
 We use the moving window technique to construct time-varying correlation matrices and PMFG networks. Considering the temporal properties of the stock market, it is impossible to fully
 describe the whole system with a single adjacency matrix. Previous studies try to resolve this problem
 by aggregating temporal networks into a static network\cite{Holme2012}. However, the obvious drawback
 of this approach is that the information about the time evolution of the system is missing. 
 Very recently, the research about temporal and multilayer network have become the new frontier of network science\cite{Kivela2014,Boccaletti2014,Taylor2015}.
 The mathematical formulation of the multilayer network provide us a possible way to describe the
  temporal network structure in a unified way. Since the only difference between temporal network and 
  multilayer network is the direction of the coupling between each layer. Thus we treat the temporal 
  stock network as a special case of multilayer network and analyze its properties based on the supra-adjacency matrix\cite{Taylor2015,DeDomenico2014a} Actually the supra-adjacency matrix concept has
  already been used to describe the temporal networks in Ref.\cite{Taylor2015,Huang2017}.
  
  Here a series of PMFG networks can be described as $G^{t}=(V,E)^{t}, t\in (1\ldots T)$. The adjacency 
  matrix of PMFG $G^{t}$ at time $t$ is denoted by $A^{t}$. For the temporal stock network, the network size $N$ of each time slice
  is fixed. The coupling matrix between different time layers is an $N\times N$ dimension matrix 
  $W_{t_{a}t_{b}}$. Then the supra-adjacency matrix with dimension $NT\times NT$ can be written as,\\
  	\[\textbf{A}=\left(
  \begin{array}{cccc}
  A^{1} & W_{12} & \cdots & W_{1T}\\
  W_{12} & A^{2} & \cdots & W_{2T}\\
  \vdots& \vdots& \ddots & \vdots\\
  W_{T1}& W_{T2}& \cdots & A^{T}
  \end{array}
  \right),\]
 here $\textbf{A}$ is the supra-adjacency matrix with bidirectional coupling. However, for temporal
 network the coupling is directional. So the upper triangle of the supra-adjacency matrix should be
 zero. As described in Ref.\cite{Huang2017}, the supra-adjacency is named as supra-evolution matrix with a 
 time directional coupling. The adjacency matrix $A^{t}$ is easy to obtain. The big challenge here is how to determine the 
 coupling matrix $W_{t_{a}t_{b}}$. The temporal stock network is different from the real
 multilayer network for which the coupling between each layer is well defined. Thus we employ the 
 time series analysis method to model the evolution of the stock network. The coupling between two 
 networks at successive time slices can be obtained from time series modeling. We use the
 autoregressive moving average model($\mathrm{ARMA}$) to fit
 the correlation strength time series of each stock. Considering the non-stationarity of the 
 correlation strength time series, before the $\mathrm{ARMA}$ model is applied, we need to difference 
 those time series to make them meet the stationary requirements meaning that the actual correlation 
 strength time series can be fitted with the $\mathrm{ARIMA}(p,d,q)$ with differencing order $d$. The $\mathrm{ARMA}(p,q)$ model is described as\cite{BG2015}:
 \begin{align*}
 s_{i,t}=\phi_{i,1}s_{i,t-1}+\phi_{i,2}s_{i,t-2}+\ldots+\phi_{i,p}s_{i,t-p}\\
 +e_{t}-\theta_{i,1}e_{t-1}-\theta_{i,2}e_{t-2}-\ldots-\theta_{i,q}e_{t,q},
 \end{align*}
where $s_{i,t}=\sum\limits_{j=1}^{N}\rho_{i,j}^{t}$ is the correlation strength of stock $i$ at time $t$. $e_t$ is Gaussian noise. Whist $\Phi_{i,p}=(\phi_{i,1},\phi_{i,2},\ldots,\phi_{i,p})$ and 
$\Theta_{i,q}=(\theta_{i,1},\theta_{i,2},\ldots,\theta_{i,q})$ are the model parameters(AR and MA parts) with model orders $p$ and $q$.\\

 The autoregressive parameters $\Phi_{i,p}$ specify that the correlation strength $s_{i,t}$ of node $i$ depends linearly 
 on its own previous $p\mathrm{th}$ values. Thus the coupling matrix $W_{t_a,t_b}$ for $t_a>t_b$ can be written as
 	\[W_{2,1}=\ldots=W_{t,t-1}=(\phi_{i,1})_{N\times N}=\left(
 \begin{array}{cccc}
 \phi_{1,1}&0&\cdots&0\\
 0&\phi_{2,1}&\cdots&0\\
 \vdots& \vdots& \ddots & \vdots\\
0&0&\cdots&\phi_{N,1}
 \end{array}
 \right)\]
 \begin{align*}
 	W_{t,t-2}&=(\phi_{i,2})_{N\times N}, i=1,2,\ldots, N,\\
 	&\ldots\\
 	W_{t,t-l}&=(\phi_{i,l})_{N\times N}, i=1,2,\ldots, N.
 \end{align*}
 While for $t_a < t_b$, we set $W_{t_a,t_b}$ to zero matrix. So the supra-evolution matrix is a lower
 triangle block matrix
 	\[\textbf{A}=\left(
 \begin{array}{cccc}
 A^{1} & 0& \cdots & 0\\
 W_{2,1} & A^{2} & \cdots &0\\
 \vdots& \vdots& \ddots & \vdots\\
 W_{T,1}& W_{T,2}& \cdots & A^{T}
 \end{array}
 \right).\]
 
 With the supra-evolution matrix, we can define some centrality
 measure to quantify the importance of different stocks. Many centrality
 measures are based on the element of leading eigenvector corresponds to 
 the largest eigenvalue of different matrices(e.g., adjacency matrix). The temporal centrality 
 can be defined by the largest eigenvalue and corresponding eigenvector
 of the supra-evolution matrix, i.e.,
 \begin{align}
 	\mathrm{\textbf{A}}\boldsymbol{\nu_1} =\lambda_1 \boldsymbol{\nu_1},
 \end{align}
 where $\boldsymbol{\nu_1}$ is the eigenvector corresponding to the largest eigenvalue $\lambda_1$ with dimension
 $NT\times 1$, $\boldsymbol{\nu_1}=(\nu_{i}^{t})_{NT\times1},i=1,2,\ldots, N; t=1,2,\ldots, T$. The element $\nu_{i}^{t}$ 
 represents the centrality value of node $i$ at time $t$. Thus for node $i$ in temporal stock
 network, the eigenvector centrality $c_i$ can be defined as the summation of the value of $\nu_{i}^{t}$
 in different time slices, namely,
 \begin{align}
 	 c_i=\sum\limits_{t=1}^{T}\nu_{i}^{t},i=1,2,\ldots,N.\label{tmpCentral}
 \end{align}
 
\section{Results and Application\label{results}}
\subsection{Topology analysis of temporal stock networks}
\begin{figure}[ht]
	\centering
	\includegraphics[width=\textwidth]{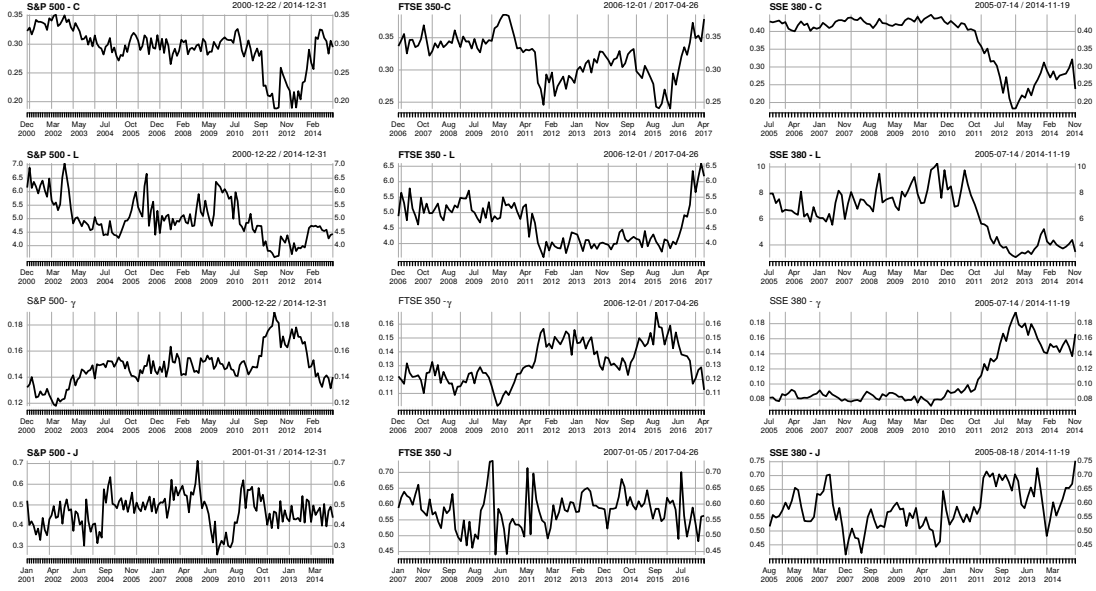}
	\caption{\label{topology}The time evolution of the topological quantities of the PMFG networks for three markets. The first column is the topological quantities of the 
	PMFG networks for S\&P 500 constituent stocks. The second column is the topological quantities of the PMFG networks for the FTSE 350 constituent stocks. The third column is the topological quantities of the PMFG networks for the SSE 380 constituent stocks. The first row is the clustering coefficient $C$ for three markets. The second row is the shortest path length $L$ for three markets. The third row is the heterogeneity index $\gamma$ for three markets. The last row is the Jaccard index $J$ for three markets.}
\end{figure}
In Fig.\ref{topology}, we show the time evolution of the topological parameters of PMFG networks for the three markets. For the US stock market, the topology structures of the PMFG networks respond to the 2008 sub-prime crisis during which the Jaccard index decreased dramatically. It means the market suffered from extremely
unstable period with drastic structure variation. For the UK market, during the European debt crisis, the
clustering coefficient $C$ and shortest path length $L$ both decreased. The heterogeneity index $H$ of the
PMFG network increase significantly during the crisis. The reaction of the correlation-based networks during financial crisis has been systematically investigated\cite{Tumminello2010,Song2011,Wang2013a,Zhao2015b,Nobi2014b,Kenett2015,Wang2015a}. Here we find that the heterogeneity index of China stock market is apparently small before 2012 with higher clustering coefficient $C$ and longer shortest path length $L$. It is known that the heterogeneity value $H$ of the scale-free network is 0.11. The western markets are more heterogeneous than the scale-free network and they are considerably more heterogeneous than China market. The homogeneous structure of Chinese market before 2012 indicates that the Chinese market has totally different structure compare to the western markets. During the period between 2011 and 2014, the Chinese stock market suffered 
from a long term bear market. The market heterogeneity increased dramatically
during that period. This means that the market try to get rid of the domination of the 
index or the market trend, which maybe resulted from the risk diversification of 
the investors or the market becoming mature. Although we can obtain some information from the variation of those
topological parameters, those quantities suffer from the very unstable market states and strong noise. The 
evolution of those topology quantities indicate that the markets are always evolving over time. The temporal 
properties of the stock markets should be considered and incorporated into real life applications. In the next
section, we try to utilize the temporal attributes to improve the performance of the portfolio optimization procedure.

\subsection{Portfolio optimization}
 \subsubsection{Mean-variance portfolio optimization}
 We first employ the PMFG networks to 
 improve the performance of portfolio optimization under the Markowitz portfolio optimization
 framework\cite{Markowitz1952}. There are lots of works trying to establish connections between the correlation-based 
 networks and the portfolio optimization problems\cite{Onnela2003b,Tola2008235,Peralta2016}. We now give an brief introduction about the Markowitz 
 portfolio theory. Consider a portfolio of $m$ stocks with return $r_i,i=1\ldots 
 m$. The return $\Pi(t)$ of the portfolio is
 \begin{align*}
 \Pi(t)=\sum\limits_{i=1}^{m}\omega_{i}r_{i}(t),
 \end{align*}
 where $\omega_{i}$ is the investment weight of stock $i$. $\omega_{i}$ is normalized such that
  $\sum\limits_{i=1}^{m}\omega_{i}=1$. The risk of the portfolio can be simply quantified by the variance of the return\\
 \begin{align*}
 \Omega^{2}=\sum\limits_{i=1}^{m}\sum\limits_{j=1}^{m}\omega_{i}\omega_{j}\rho_{ij}\sigma_i\sigma_j,
 \end{align*}
 here $\rho_{ij}$ is the Pearson cross-correlation between $r_i$ and $r_j$, and $\sigma_i$ and $\sigma_j$
 are the standard deviations of the return time series $r_i$ and $r_j$. The optimal portfolio weights are determined via maximize the portfolio return 
 $\Phi=\sum\limits_{t=1}^{T} \Pi(t)$ under the constraint
 that the risk of the portfolio equals to some fixed value $\Omega^{2}$. Maximizing $\Phi$ 
 subject to those constraints above can be formulated as a quadratic 
 optimization problem:
 \begin{align*}
 \omega^{T}\Sigma\omega-q*\mathrm{R}^{T}\omega,
 \end{align*}
 where $\Sigma$ is the covariance matrix of the return time series. The parameter $q$ is the risk 
 tolerance parameter with $q\in [0,\infty)$. Large $q$ indicates that the investors have strong tolerance to the risk
  which may give large expected return. Whilst, small $q$ represents that the investors are extremely risk aversion.
 The optimal portfolios at different risk and return levels can be presented as 
 the efficient frontier which is a plot of the return $\Phi$ as a function 
 of risk $\Omega^2$. 
 
 So far we have not illustrate how to determined the constitute stocks of a specific portfolio. 
 As mentioned in the previous context, we use some centrality metric to choose portfolio from the 
 PMFG networks. It has shown that the performance of the portfolio selected 
 by using some compound centrality measures for the static PMFG networks is quite good\cite{Pozzi2013,Zhao2017a}.
 Here we try to select the portfolio guided by the temporal eigenvector centrality measure of the temporal
 PMFG networks for different stock markets. A portfolio constructed by using the central
 (peripheral) stocks is the one that consists of those higher (lower) centrality value stocks. For comparison, we also perform the 
 portfolio optimization procedure based on aggregated networks\cite{Holme2012}. For the aggregated network, 
 we use the compound centrality measure from Ref.\cite{Pozzi2013} to rank the stocks. In contrast, in the temporal 
 stock networks, the stocks are ranked according to the temporal centrality given by Eq. \ref{tmpCentral}.
 To verify the robustness of the portfolios' performances, we performed both in sample and out of sample 
 tests for those temporal portfolios.
   \begin{figure}[ht]
  	\centering
  	\includegraphics[width=\textwidth]{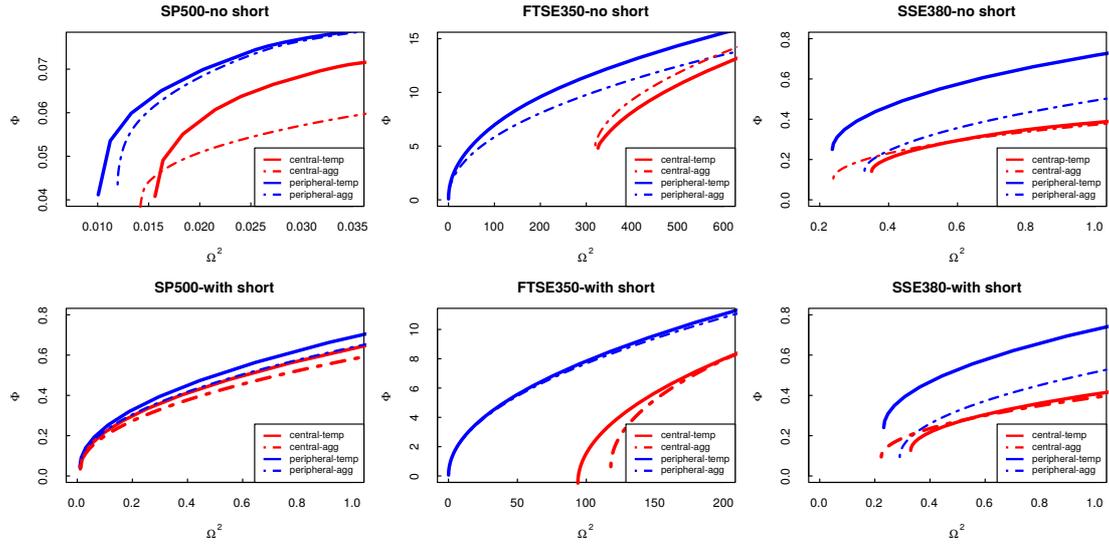}
  	\caption{\label{eff}The in sample efficient frontiers for three different stock markets. The left, center and right columns are the results for
  		 S\&P 500, FTSE 350 and SSE 380 respectively. The red lines are the results for those portfolios constructed from stocks with high centrality
  	 scores(central) for both temporal(suffix -temp) and aggregated(suffix -agg) networks. The blue lines are the results for those portfolios constructed from stocks with
   low centrality scores(peripheral). Here the portfolio size $m=30$. We have tested the portfolio size from $m=5$ up to $m=60$, the results are consistent.}
  \end{figure}

Fig. \ref{eff} shows the in sample efficient frontiers of a portfolio constructed by those stocks with 30 highest centrality
 and 30 lowest centrality stocks for both aggregated and temporal stock networks. Here during the in sample tests, the whole
 datasets(with 4205, 3000 and 2700 records for US, UK and China respectively) have been used to
 construct the temporal networks and the portfolio optimization is also performed with the whole datasets. The solid lines are those portfolios selected guided by the eigenvector centrality 
 for temporal PMFG networks. The dashed lines are those portfolios for aggregated networks. 
 The aggregated network is constructed by combining all the vertices and edges in all the time slices of temporal networks.
 The solid and dashed red (blue) lines are those portfolios of central(peripheral) stocks. It is very clear
 that the performance of the peripheral portfolios are much better than those central ones for three markets. 
 That is exactly in line with the previous research.
 Meanwhile, the in sample performance of portfolios for temporal networks(solid lines) are also better than those constructed from 
 aggregate networks(dashed lines). The overall best in sample performance comes from those portfolios constructed based on temporal networks and peripheral stocks(solid blue lines). Those portfolios have the highest return and the lowest risk compared with other portfolios.
 \begin{figure}
 	\centering
 	\includegraphics[width=\textwidth]{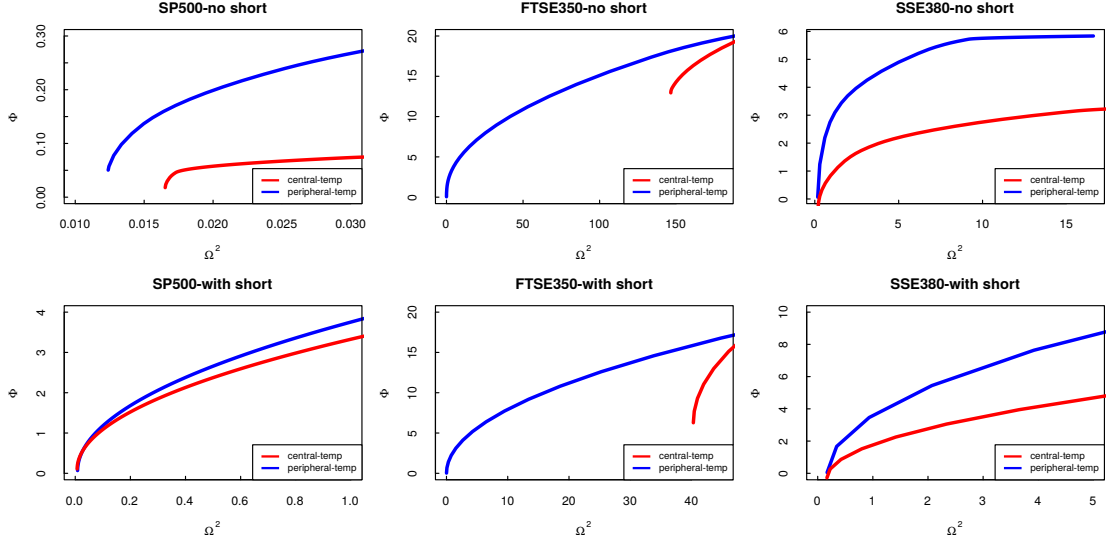}
 	\caption{\label{eff-out}The out of sample efficient frontiers of three different stock markets. The left, center and right columns are the results for
 		SP500, FTSE350 and SSE380, respectively. The red lines are the results for those portfolios constructed from stocks with high centrality
 		scores(central) for both temporal(suffix -temp) and aggregated(suffix -agg) networks. The blue lines are the results for those portfolios constructed from stocks with
 		low centrality scores(peripheral). Here the portfolio size $m=30$. We have tested the portfolio size from $m=5$ up to $m=60$, the results are consistent.}
 \end{figure}

 The out of sample tests are also performed to check the robustness of the temporal network portfolios.
 Here in Fig. \ref{eff-out}, we perform the out of sample tests for temporal portfolios. First we use the first 3500, 1650 and 1500 data points for US, UK and China markets to construct the temporal networks.
 With the guidance of temporal centrality, we can construct the central and peripheral portfolios. Then the next 225 data points are used to perform the portfolio optimization procedure. The results 
 are very similar to the in sample tests. The temporal peripheral portfolios have consistent good performance
 over the central portfolios. A very interesting phenomena is that the central portfolios of UK market always
 have very high risk for both in sample and out of sample tests. The central portfolios can not attain risks 
 lower than some specific level even for very small risk tolerance parameter $q$. This implies that the central
 stocks of UK market are extremely risky which should definitely be avoided by investors.
 The above portfolio optimization results evidence the usefulness of temporal centrality metric. The temporal information of the correlation-based networks should be taken
 into consideration when dealing with time evolving systems.

 \subsubsection{Expected shortfall approach}
 Apart from the mean-variance framework, the expected shortfall($ES$) is a more modern tool of quantifying the performance of a portfolio, which is a coherent risk measure\cite{Acerbi22001,Caccioli2013,CACCIOLI2016}. Let $X$ be the profit loss of a portfolio within a specified time
 horizon $(0, T)$ and let $\alpha = \eta\% \in (0, 1)$ be some specified probability level. The expected $\eta\%$ shortfall
 of the portfolio can be defined as
  \begin{align}
  ES^{\alpha}(X)=-\frac{1}{\alpha}(E[X\textbf{1}_{X\leq x^{\alpha}}]-x^{\alpha}(\mathrm{P}[X\leq x^{\alpha}]-\alpha)).
  \end{align}
  The $ES$ gives the expected loss incurred in the $\eta\%$ worst situations of the portfolio. It satisfies all the requirements of a risk measure.
  For a portfolio $\{\omega_{i},i=1,\ldots,m\}$ of $m$ stocks with return $\{r_i,i=1,\ldots 
 ,m\}$, we want to minimize the $ES^{\alpha}$ of the portfolio under the constraint of $\sum\limits_{i=1}^{m}\omega_{i}=1$. Here we set the confidence level $\alpha=95\%$ for the expected shortfall $ES^{\alpha}$ of the portfolio and assume that the short selling is prohibited. After ranking the stocks according to the centrality 
 scores described in the previous subsection, we choose the portfolio size $m=5,10,\ldots,55, 60$, namely, $m$ 
 central(peripheral) stocks with the largest(smallest) centrality scores.
 \begin{figure}[ht]
 	\centering
 	\includegraphics[width=\textwidth]{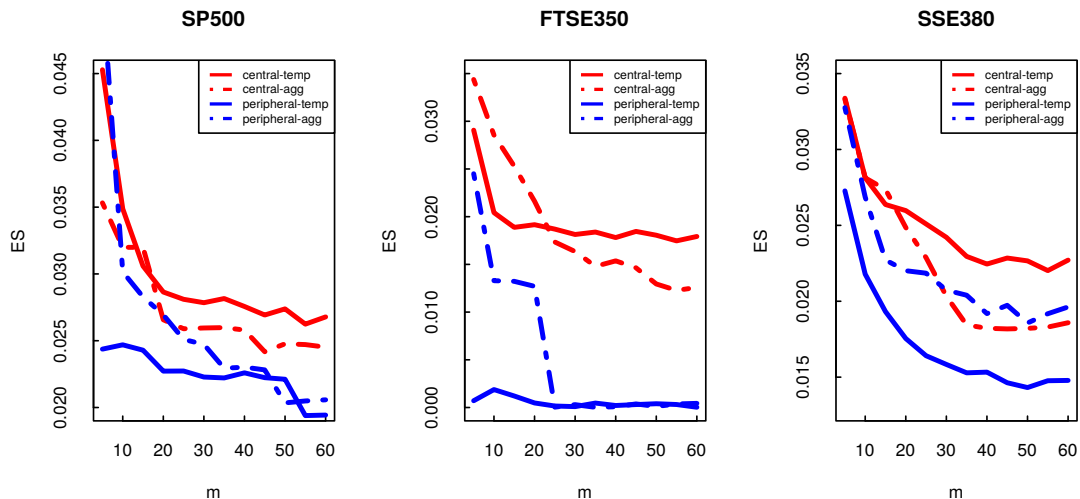}
 	\caption{\label{ES}The in sample expected shortfalls for three stock markets. The left, center and right columns are the results for
 		S\&P 500, FTSE 350 and SSE 380, respectively. The red(blue) lines are the expected shortfalls for the portfolios constructed by central(peripheral) stocks. The solid lines are the expected shortfalls for temporal networks. The dashed lines are those for aggregated(suffix -agg) networks and the solid liens are those for temporal(suffix -temp) networks.}
 \end{figure}

Fig .\ref{ES} gives the in sample expected shortfalls for the three stock markets. The red(blue) lines represent the expected shortfalls for central(peripheral) portfolios. The solid(dashed) lines corresponds to the temporal(aggregated) networks. It is obvious that the expected shortfalls for peripheral portfolios are much smaller than the central ones. An argument has been given in Ref.\cite{Peralta2016} in which the correlation matrix can be recognized as an weighted
fully connected network. There exists a negative correlation between the weights of the optimal portfolio  and the network's eigenvector centralities.
The lower expected shortfalls of peripheral portfolios have verified this argument. Whilst,
the temporal centrality as a portfolio selection tool performs even better than the static aggregated 
network centrality up to $m=60$ portfolio size. 
\begin{figure}[ht]
	\centering
	\includegraphics[width=\textwidth]{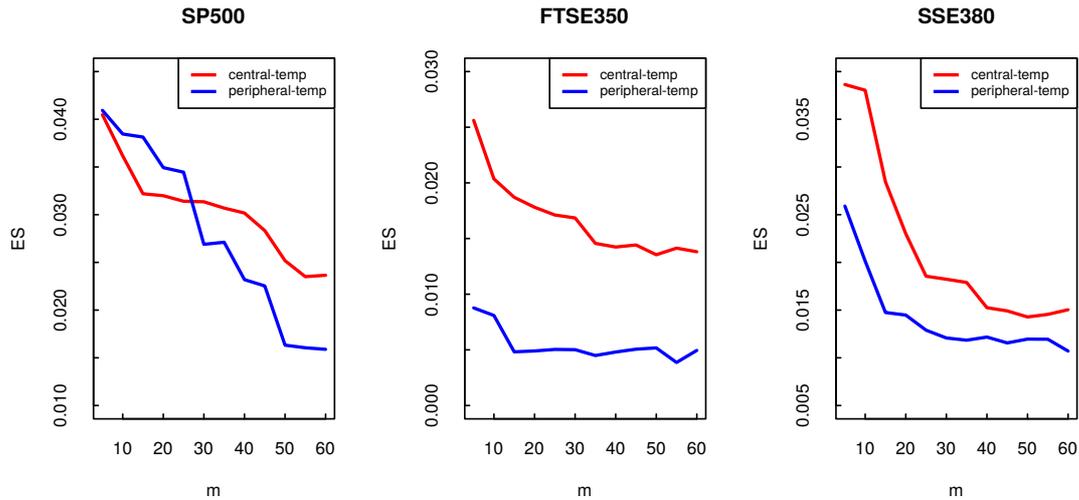}
	\caption{\label{ES-out}The out of sample expected shortfalls for three stock markets. The left, center and right columns are the results for S\&P 500, FTSE 350 and SSE 380, respectively. The red(blue) lines are the expected shortfalls for the portfolios constructed by central(peripheral) stocks.}
\end{figure}
In Fig. \ref{ES-out}, the out of sample tests are also performed for three markets. The datasets used
for the out of sample tests are exactly the same as in previous subsection. Except for the temporal 
portfolio with size $m=5$ of the US market, the peripheral portfolios for the three markets with portfolio size
up to $m=60$ all have better performances with lower expected shortfalls. We argue that the consistent good 
performance of the temporal portfolio rooted in the time average attribute of the temporal centrality. It 
can weaken the influence of large fluctuations of the market, thus it can be used to construct more robust 
and risk diversified portfolio\cite{Filiasi,Schafer2010,Zhao2017a}.
 \section{Conclusion\label{conclusion}}
In conclusion, we have used the temporal network scheme to analyze the temporal evolution of three major stock markets. The topology evolution of the correlation-based networks for three markets give some signals of corresponding financial turbulences in each market. With the help of temporal centrality measure, we can construct some risk diversified portfolios with high return and low risk. Under both the mean-variance and expected shortfall frameworks, the portfolios constructed with those peripheral stocks in both temporal and static centrality measures outperform those portfolios constructed with central stocks. Moreover, those peripheral portfolios selected with the guidance of
temporal centrality measure performed way better than other portfolios(central portfolios and aggregated 
peripheral portfolios) under both mean-variance and expected shortfall evaluation scheme. The in sample and 
out of sample tests have verified the robustness of the temporal peripheral portfolios. This is the first study 
to analyze the time evolving correlation-based networks with temporal network theory. The application of temporal 
centrality measure on portfolio selection has revealed the importance of the temporal attributes of the 
correlation-based networks of stock markets. Thus it should be quite interesting to investigate the temporal structure of the 
correlation-based networks with other tools developed for temporal network\cite{Holme2012}. This should be subject to future research.
\section{Acknowledgment}
This work is supported in part by the Programme of Introducing Talents of Discipline
to Universities (Grant no.~B08033), the NSFC (Grant no.~71501066), the Hunan
Provincial Natural Science Foundation of China (Grant no.~2017JJ3024), and the program of China Scholarship Council (Grant no.~201606770023).


 \end{document}